# Agile Transformation: A Summary and Research Agenda from the First International Workshop

Leonor Barroca[1], Torgeir Dingsøyr[2], Marius Mikalsen[2]

[1]The Open University, UK
[2]SINTEF, Norway
`leonor.barroca@open.ac.uk`

**Abstract.** Organisations are up-scaling their use of agile. Agile ways of working are used in larger projects and also in organisational units outside IT. This paper reports on the results of the first international workshop on agile transformation, which aimed to focus research on practice in a field which currently receives great attention. We report on participants' definitions of agile transformation, summaries of experiences from such transformations, and the challenges that require research attention.

## 1  Introduction

In order to increase their ability to sense, respond and learn, organisations are up-scaling their use of agile. This implies that agile is used not only in larger projects and programs, but also in other organisational units outside of IT. In a foreword to the book "Unlocking Agility" [1], Bjarte Bogsnes writes: *"The agile mindset is now finding its way into the C-suite, and it is starting to radically change the way organizations are led and managed. Business agility is on everybody's lips, for very good reasons"*.

While the implementation of agile methods traditionally has been studied at team level ([2], [3]), adopting agile practices across the organisation is widening this perspective and has been labelled "agile transformation". Research has discussed three main areas of such transformations. First, challenges and success factors in the transformation process ([4], [5], [6], [7], [8], [9],[10]); second, changes in roles and practices that occur during such transformations ([11], [12],[13]); and third, models for understanding agile transformations ([14], [15]). As an emerging research field, there are many understandings of what agile transformation is; also, current empirical studies tend to be descriptive and place little emphasis on theory to explain findings. This was the motivation to host the first international workshop on agile transformation in order to focus research on practice in a field which receives great attention.

This paper summarises the workshop, which was conducted in half a day at the *International Conference on Agile Software Development*, XP 2019. The goal of the workshop was to challenge the scientific community to identify what should be of prime interest to researchers in the area of agile transformations, as there are growing opportunities to study them as companies increasingly adopt agile. Organisations are learning from agile practice to embrace agility in their ways of working; agile practitioners can also benefit from the wider context of organisations undergoing agile transformations, to understand their wider implications, and how to sustain them. The workshop received six submissions out of which four were selected for presentation. Maria Paasivaara was invited to give a keynote talk on tips for successful agile transformations. Following the presentations, participants offered definitions of agile transformation and discussed, in an open space format, the main research challenges in this area.

The remainder of this paper reports the results of the workshop, and is structured as follows. Section 2 presents the definitions of large-scale agile transformation from participants. Section 3 provides an outline of the four presentations and of the keynote. Section 4 provides an overview of key research challenges identified by the participants at the workshop and at a similar workshop in London. Section 5 concludes the paper.

## 2      What is an agile transformation?

For many organisations moving towards business agility is challenging as there are many elements at play, from culture and leadership to process and tools. The participants in the workshop proposed different definitions for agile transformation as shown below; the terms 'culture', 'reactive/responsiveness to change' and 'continuous improvement' figured in several of them.

**Table 1.** Some of the definitions of agile transformation gathered at the workshop.

| |
|---|
| "an individual's, team's, group's and organisation's journey into continuous improvements changing the way we do business, meet our goals and overcome our challenges by being more flexible, targeting smaller goals and providing continuous delivery, feedback and learning the process which evolves an organisation to be more reactive to changes in its environment" |
| "digital transformation -> agile becomes larger (programs, portfolios) and more important; also becomes more complex, needs alignment with other units that are not traditionally agile; change in leadership and management" |
| "a people-centred approach to improving business outputs in the context of its environment the process undertaken to develop capabilities that will allow for flexibility in responding to a changing environment and continuous improvement" |
| "a path from adopting agile practices to establishing agile culture" |
| "transform from rather rigid structures, processes and hierarchy to a more network organisation with increased knowledge, understanding and collaboration across boundaries to improve a company's reaction to external change in order to improve performance referring to effectiveness" |
| "shift towards practices that enable organisational responsiveness" |
| "agile – iterative, incremental, collaborative, effects/results/outcomes-driven transformation – continuous improvement from where you are towards the Agile values and principles" |

## 3      Experience with Agile Transformation

Lucas Green presented an industry case study of a big bang transformation with processes as usual having to coexist with new processes and resulting challenges; a research-based questionnaire was used to help understand team maturity during the transformation. A lesson from this case study is that the key to obtain understanding during the transformation is to support self-organising teams.

Akim Berkani discussed agile transformation beyond IT based on a case study of a French administration department. His approach took a management innovation implementation lens (new structures, practices and processes) to explain the transformation process.

Johannes Berglind, Ludvig Lindlöf, Lars Trygg and Rashina Hoda presented a study of an agile transformation in the automotive industry; their study was conducted bottom-up with the engineers who already practiced agile informally before the top-down transformation was carried out. They highlighted the paradox of top-down transformations not taking into account the informal agile practices already in place. They suggested an approach that takes into account these informal agile practices already present incorporating them into the transformation.

Katie Taylor took the lens of a practitioner business agility framework (www.agile-business.org) to identify the leadership elements needed for an agile transformation. She proposed the analogy of 'head, heart and hands' therefore focusing on people and their central role in agile transformation.

Maria Paasivaara gave the keynote at the workshop on tips for a successful agile transformation. They were based on past and ongoing studies of transformation processes in more than six private companies from sectors such as telecom and finance, and also from the public sector. The tips included ensuring management support, involving everyone in the organisation, communicating reasons for change, training everyone, hiring coaches with experience to help the transformation, ensuring transparency, developing an agile mindset, customising the transformation to fit organisation and product, including effort to improve, focusing on systems thinking as well as physical workspace and infrastructure.

## 4      Research agenda on main challenges

We carried out a survey with the workshop participants; the same survey had been carried out with the participants of a similar workshop targeted at industry that took place in London, UK, two weeks before. The total of 39 respondents were distributed as shown in Figure 1, with the majority having management positions (35%), followed by research scientists (29%) and agile coaches (18%). The 'Other' group was composed of two consultants and one software architect. We do not claim that the respondents to this informal survey are in any way a representative sample of software companies or researchers in software engineering, but they are people interested enough in the topic to devote half a day to discuss it; they had an average of 6 years of experience with agile transformations (standard deviation: 6).

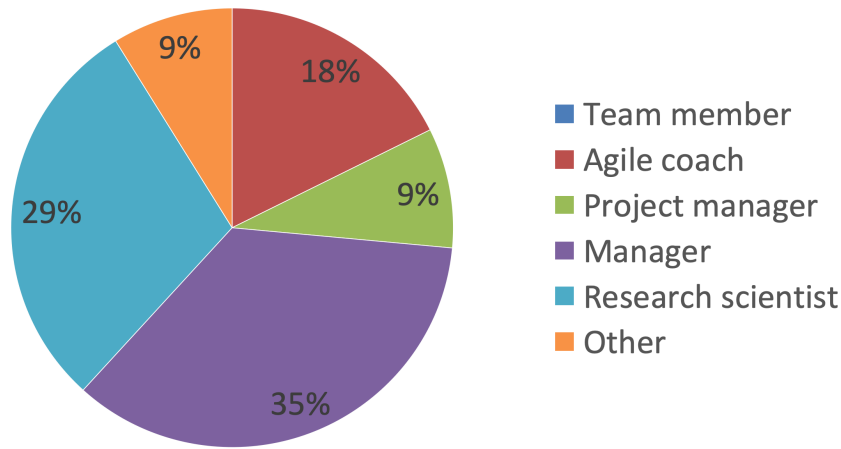

**Fig. 1.** *Roles of participants at the workshops who completed the survey.*

Participants ranked their motivation for agile transformation after a scale taken from the state of agile survey[1]; the top three reasons were: 'improve business/IT alignment', 'enhance ability to managing changing priorities' and 'accelerate software delivery'.

Participants were also asked to rank success factors and challenges slightly modified from [4], based on own knowledge of transformation projects. They ranked the top three success factors to be: 'changing organisational culture', 'leadership' and 'engaging people'.

Challenges were ranked as shown in Table 2. Participants could add 'Other' challenges to the list, which resulted in three more challenges: 'shareholder value dominates operating models', 'competence-building and empowerment of teams' and 'operations'. One respondent answered that the challenges available were 'not good'.

**Table 2.** Ranked challenges in Agile Transformations, challenges taken from [4].

|   | Challenge | Description |
|---|---|---|
| 1 | Hierarchical management and organizational boundaries | Middle managers' role in agile unclear<br>Management in waterfall mode<br>Keeping the old bureaucracy<br>Internal silos kept |
| 2 | Integrating non-development functions | Other functions unwilling to change<br>Challenges in adjusting to incremental delivery pace<br>Challenges in adjusting product launch activities<br>Rewarding model not teamwork centric |
| 3 | Resistance to change | General resistance to change<br>Scepticism towards the new way of working<br>Top down mandate creates resistance |

---

[1] See VersionOne state of agile report: https://www.stateofagile.com

| | | Management unwilling to change |
|---|---|---|
| 4 | Coordination challenges in multi-team environment | Interfacing between teams difficult<br>Autonomous team model challenging<br>Global distribution challenges<br>Achieving technical consistency |
| 5 | Agile difficult to implement | Misunderstanding of agile concepts<br>Lack of guidance from literature<br>Agile customised poorly<br>Reverting to old ways of working<br>Excessive enthusiasm |
| 6 | Lack of investment | Lack of coaching<br>Lack of training<br>Too high workload<br>Old commitments kept<br>Challenges in rearranging physical work space |
| 7 | Different approaches emerge in a multi-team environment | Interpretation of agile differs between teams<br>Using old and new approaches side by side |
| 8 | Quality assurance challenges | Accommodating non-functional testing<br>Lack of automated testing<br>Requirements ambiguity affects QA |
| 9 | Requirements engineering challenges | High-level requirements management largely missing in agile<br>Requirement refinement challenging<br>Creating and estimating user stories hard<br>Gap between long and short term planning |

In both workshops, we discussed four of the main challenges and tried to identify relevant theory and research methods for future studies on these topics. More detailed minutes are available in a summary at the XP2019 programme website.

## 5  Conclusion

This workshop showed that the research community is interested in continuing studies on agile transformations, and that there is a growing body of studies on which to build up. We hope the initial research agenda developed at the workshop will inspire future studies.

## 6  Acknowledgement

Thanks to all presenters and participants and to Rashina Hoda for chairing the research workshops. Further, we are very grateful to the program committee members: Finn Olav Bjørnson (Norwegian University of Science and Technology), Andreas Drechsler (Victoria University of Wellington, New Zealand), Peggy Gregory (University of Central Lancashire, United Kingdom), Lucas Gren (Volvo Cars, Sweden), Tomas Gustavsson (Karlstad University, Sweden), Teemu Karvonen (University of Oulu,

Finland), Per Lenberg (Saab AB, Sweden), Will Menner (Johns Hopkins University, USA), Shannon Ewan (ICAgile, USA), Kai Spohrer (University of Mannheim, Germany), Helen Sharp (Open University, United Kingdom), Katie Taylor (Agile Business Consortium, United Kingdom), Marianne Worren (Norwegian Labour and Welfare Administration).